\documentclass{article}

\usepackage{arxiv}

\usepackage[utf8]{inputenc} 
\usepackage[T1]{fontenc}    
\usepackage{url}            
\usepackage{booktabs}       
\usepackage{amsfonts}       
\usepackage{graphicx}
\usepackage{nicefrac}       
\usepackage{microtype}      
\usepackage{lipsum}
\usepackage{hyperref}  
\title{
Theta, alpha and gamma traveling waves in a multi-item working memory model
}

\author{
  Gustavo D. Soroka\\
  Neuroscience Program\\
  Federal University of Rio Grande do Sul\\
  Porto Alegre, RS \\
   \And
 Marco A. P. Idiart\\
  Physics Department\\
  Federal University of Rio Grande do Sul\\
  Porto Alegre, RS 
}

\begin{document}
\maketitle

\begin{abstract}
Brain oscillations are believed to be involved in the different operations necessary to manipulate information during working memory tasks. We propose a modular variant of the Lisman-Idiart model for short-term memory, where theta, alpha and gamma oscillations are traveling waves. Using this model we show that the interactions between theta and gamma determine the allocation of multiple information in distinct modules, while the interference between theta and alpha disrupts the maintenance of current stored information. The  effect of alpha in erasing or blocking storage is robust and seems fairly independent of the frequency, as long as it stays within the alpha range. The model help us to understand why the alpha and theta oscillations, which have close frequency bands, could have opposite roles in working memory.
\end{abstract}

\keywords{Working Memory \and Brain Oscillations \and Travelling Waves \and Modules \and Computational Neuroscience}

\section{Introduction}
Working memory is a putative memory system that incorporates many short-term information storage subsystems 
\cite{baddeley_1986,baddeleyehitch_1974}, and serves as  an interface between perception, long term memory and action \cite{baddeley_2007}. In doing so, it  contributes to higher cognitive functions such as reasoning, planning, decision-making and language comprehension \cite{funahashi_2006}.

To date, the most comprehensive model for a working memory system started by the work of  Baddeley and Hitch \cite{baddeleyehitch_1974}, and was later improved in a series of papers \cite{baddeley_1988,baddeley_2000}.
The model proposes different components, or subsystems, to deal with different kinds of information and processes, such as the visuospatial sketchpad for visual and spatial information, the phonological loop for auditory and phonological information, the episodic buffer for interactions between different information and a scene, and a central executive for a top-down control over the subsystems. In special, the phonological loop that could store multiple serial information was the most well-studied component for its relations with language and the experimental data accumulated over the years \cite{baddeley_1966a,baddeley_1966b,baddeley_1984,BADDELEY1975575}.

There are many proposed models for the short-term storage devices that compose the working memory system. Most of them 
agree that some sort of persistent activity of the neurons involved in storing the information is necessary, 
but disagree on the underlying physiological mechanism. The models can be divided in roughly two categories the ones that favour network mechanisms (e.g. instantaneous attractors due to shot-term synaptic changes and  continuous attractors or bump models ) and models that favour single cell phenomena .  

In the last category is the model proposed by Lisman and Idiart \cite{LismanIdiart_1995}. According to this model, incoming information of a sequence of items to be memorized causes the ordered activation of item specific neural ensembles. The firing of these cells changes their intrinsic membrane properties causing an afterdepolarization that effectively produces a  transitory excitability that peaks between 100ms and 200ms after firing. In the absence of external inputs these tagged cells can be reactivated by an  unspecific (not informational) oscillatory input with frequency in the theta range. 
It is important that the excitability have an inverted U-shaped time course so different ensembles fire in specific phases of the slow input oscillation. Another specific requirement of the model is feedback inhibition that prevent the synchronization of the different ensembles, so the
 individual memories are stored in time multiplexed fashion. A prediction of this model is a phase-amplitude coupling between the slow oscillation in the theta range representing the maintenance signal and the fast oscillations in the gamma range representing the firing of the stored memories. Subsequent studies have explored the theta-gamma mechanism in different models \cite{jensen_1996, Jensen_1998}, and a good theoretical review can be found in \cite{thetagammaneuralcode}.

Besides theta and gamma, another oscillation has been associated with an active role in cognitive processes, including working memory. This oscillation happens to be the first rhythm that was observed in humans, the alpha rhythm, almost a century ago by Hans Berger \cite{berger1929}. The increase in alpha power was related with the inhibited activity of areas during 
working memory maintenance tasks (see \cite{Uhlhaas2014} for a review). 
During working memory scanning tasks, where the subject needs to hold several items in memory for a brief period of time and then test, it was observed a increase in alpha power during the retention period and a decrease during the retrieval of information \cite{KLIMESCH1999493,Jensen2002,SCHACK2002107}. It was also observed an increase in alpha power related to the number of stored items \cite{KLIMESCH1999493,Jensen2002,SCHACK2002107}. 

A few theories have been used to describe possible roles for the alpha oscillation. The inhibition-timing hypothesis proposes that the peaks of alpha constrains time-windows of opportunities for the firing of neurons \cite{KLIMESCH200763}, with the increase of  alpha power leading to a decrease of the windows size. The gating by inhibition hypothesis supposes that alpha can select the most relevant information to be processed through the blocking of irrelevant information routes \cite{Jensen2010gating}. Being slightly different, the oscillatory selection hypothesis suggest that the information selection could be accomplish by an entrainment between sensory stimulus and brain oscillations \cite{schroeder2008}. In general, the main functional role addressed for alpha oscillations relates to the inhibition of irrelevant information.

Dippopa et. al.\cite{Dipoppa2013,Dipoppa2016} proposed a computational model where brain oscillations act as functional operators of a working memory network, in other words, applying external oscillatory currents to the network allows it to store, maintain, prevent upload and erase information . In the model the alpha oscillation is responsible for erasing the content and preventing new content of been uploaded to the network, while theta  oscillation blocks upload but maintain the current content. 

In this paper we explore theoretically the possible inhibitory role of an oscillation in the alpha range in the context of a 
 Lisman-Idiart like model. More specifically, we want to explore if the increase in alpha power in the unspecific oscillatory input can disrupt the model's multiple-item maintenance and if there is any clues of by which mechanism this may happen.
For that we introduce a modular Lisman-Idiart memory model where we introduce a spatial dimension in a form of few discrete modules and we consider that the unspecific oscillatory input, responsible in the model for the maintenance of firing, sweeps the network probing the different modules at different times as a traveling wave.

In the following sections we introduce the model, discuss the results  and  conclude. The model's mathematical description  and the computational experiments technical details  are left to the end in the Materials and Methods section.

\section{The Model}
\label{sec:headings}

\begin{figure}[ht]
\begin{centering}
	\includegraphics[width=13cm]{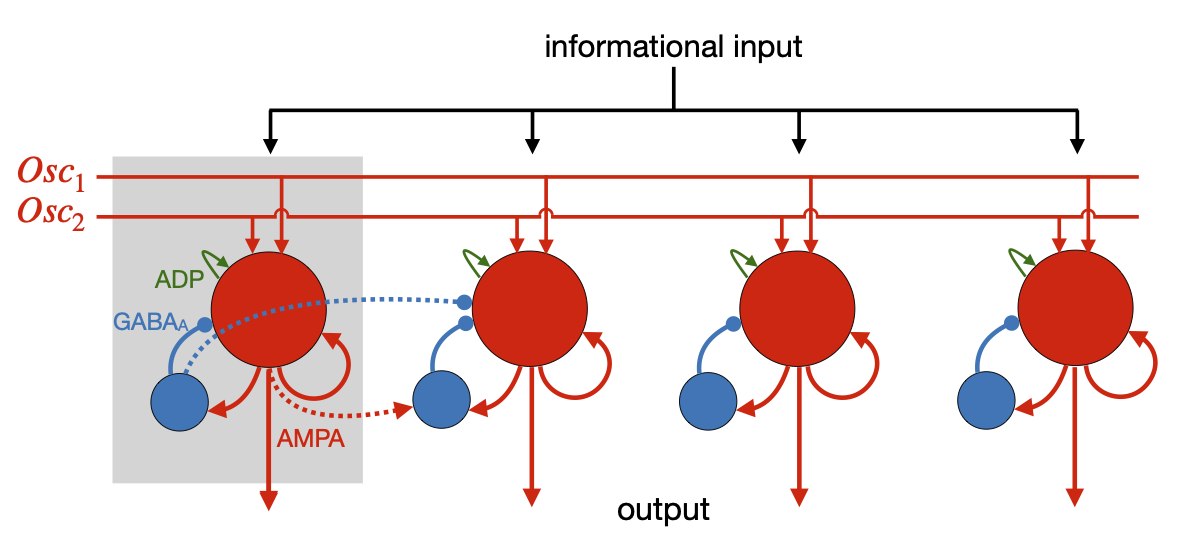} 
	\caption{Modular working memory network. The modules are arranged linearly, each one composed by excitatory principal neurons (red circles) and inhibitory interneurons (blue circles). The informational input reach all modules synchronously while 
	the unspecific oscillatory inputs ($Osc_1$ and $Osc_2$ ) are traveling waves and therefore, at a given time, the modules have oscillations with different phases.
	When in storage mode (maintain) a  spatial firing pattern is reactivated for each module, producing the cyclic phenomena observed in the rasterplot of Fig.\ref{fig22}.  } 
	\label{fig1} 
    \end{centering}
\end{figure}

Here we present  a modular version of the  Lisman-Idiart theta-gamma model. The idea is to use space to improve the robustness of the model to noise as well as to enrich the possibilities of memory storage and manipulation. For simplicity we consider that the modules are distributed in a linear fashion (see Fig.\ref{fig1}). In each module a short-term theta-gamma memory model is implemented as a network of excitatory neurons and inhibitory interneurons, that  can perform  multiple information storage through the cyclic reactivation of spatial firing patterns due to the combination of depolarizing currents after an action potential (ADP) and the excitatory drive by an unspecific  oscillatory input on the theta-alpha frequency range (6-12Hz), as the original model \cite{LismanIdiart_1995} (See Materials and Methods). 

In addition we consider that the oscillatory inputs are traveling waves in such way that each module receives an oscillation with slightly different phase.
The evidences of travelling waves in the brain date almost the first human EEG recording (see \cite{burkit_2000}), but its importance has been recently highlighted by experimental work demonstrating that theta propagates in hippocampus \cite{theta_w,theta_hip,theta_hip2}, and theta and alpha propagate in neocortex \cite{theta_alpha_cortex}. For a theoretical review, see \cite{review_sej}.

As in the original Lisman-Idiart Model, each module is capable of coding a number of overlapping firing patterns. 
But the modular version allows different items to be simultaneously stored in distinct modules that are scanned periodically by the travelling wave. It combines spatial and temporal properties such that the change in the oscillatory frequency input allows a much richer repertoire  of operations for manipulating the information held in working memory.

Dipoppa et al \cite{Dipoppa2013} proposed four essential operations for a short-term memory model: load. maintain, block and erase.  In this paper we show how each one of these operations can be performed in the context of our model, with the exception of the block operation that is not implemented independently but in association to the erase operation. 
 

 In other words, we consider that a network is blocked when it cannot hold information. The theta frequency will be responsible for allowing the load and maintain operations to take place and the alpha oscillations will perform the block/erase operation by  interfering with the theta oscillations.

\section{Results}

\begin{figure}[ht]
	\hspace{0.2cm}\includegraphics[width=1.0\linewidth]{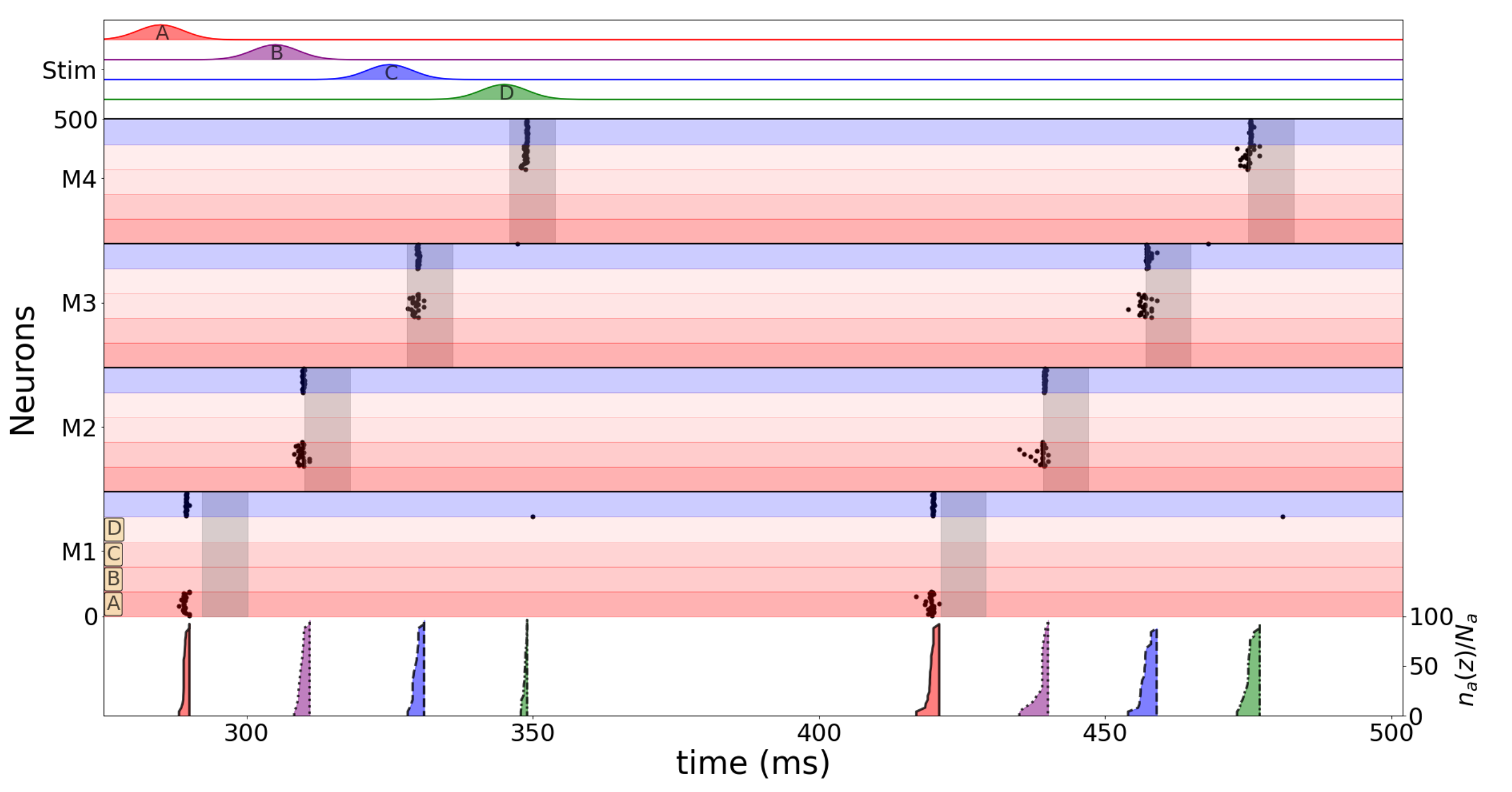} 
	\caption{Sequential allocation of items in different modules. On top: four Gaussian stimuli colored and labeled red for A, purple for B, blue for C and green for D. Middle: rasterplot of modules M1, M2, M3 and M4 during two cycles. The excitatory neurons are shown with a red gradient discriminating neurons that codes different stimuli (A-D). Inhibitory neurons are shown in blue. Shaded grey areas indicates the time windows of max neuronal excitability due to the theta wave of (>0.9 max). Bottom: cumulative frequency of the firing of neurons for each module (line style) and each stimulus (color). Label on the right, where a $\in$ [A,B,C,D]. Parameters used: $\psi_{osc}$ = 0.9 rad/module, $f_\gamma$ = 50 Hz, $f_\theta$ = 8 Hz, $\phi_i$ = 0.8 rad} 
	\label{fig22} 
\end{figure}

Figure \ref{fig22} illustrates how the  load operation, that   allocates exactly one information per module, is accomplished when sequential stimuli is fed to the network. On top we have the input currents  of the four stimuli. The currents are Gaussian shaped to account for small asynchronies from the upstream inputs.  The modules  activities are represented below with raster plots. The excitatory neurons are shown with a red gradient discriminating neurons that codes different stimuli (A-D). Inhibitory neurons are shown in blue. On the bottom, the cumulative frequency for the firing of neurons representing each stimulus for each module. The color code distinguish each stimulus and the line style distinguish the modules. Shaded areas represent the peaks of the travelling wave across the network. 

\begin{figure}[ht]
\begin{centering}
	\hspace{0.2cm}\includegraphics[width=0.7\linewidth]{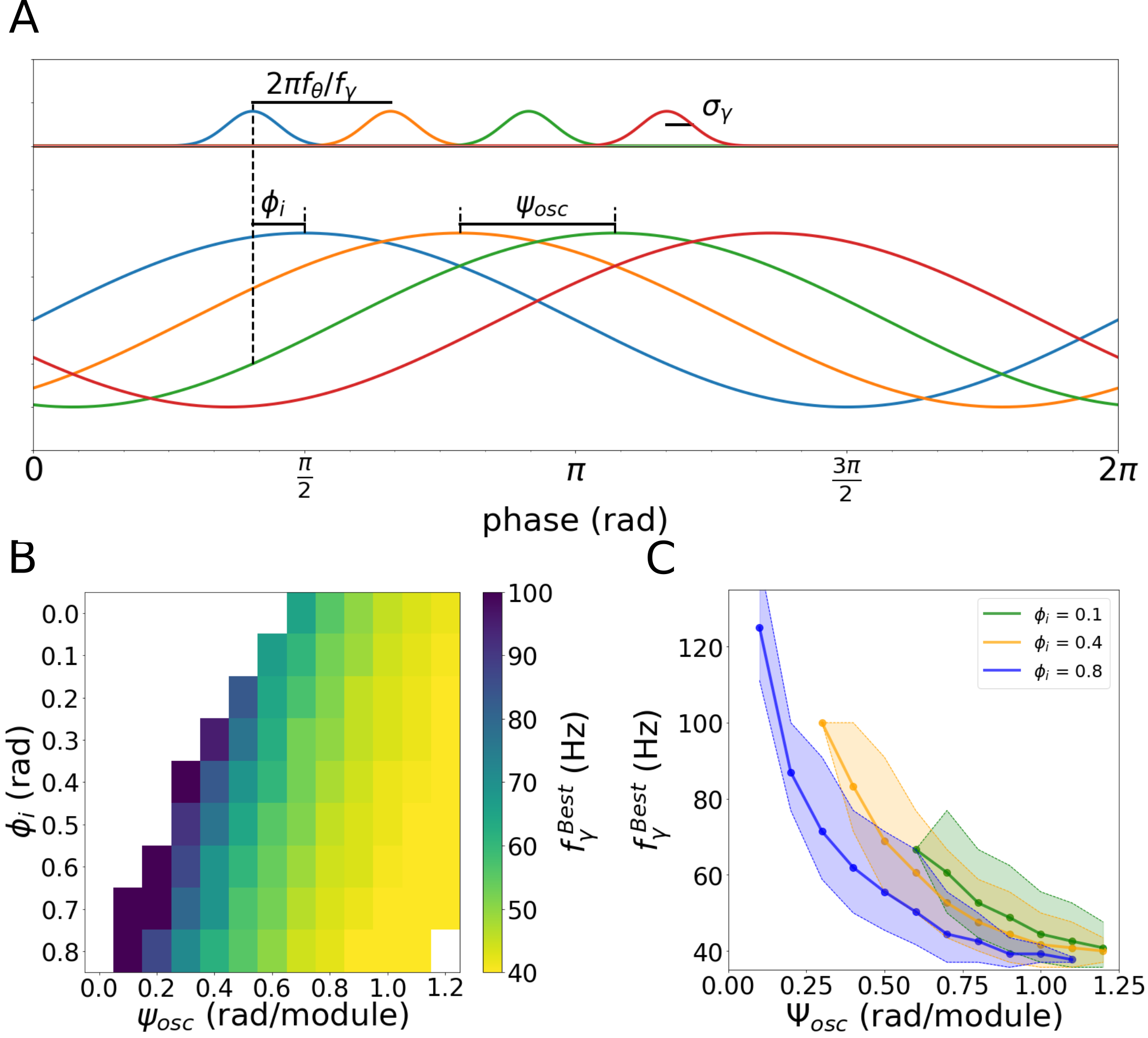} 
	\caption{Memory allocation on different modules. A) Definition of $f_\gamma$, $\psi_{\mbox{\tiny osc}}$ and $\phi_i$. Four different stimulus are given equally to all four modules, with a given $f_\gamma$ frequency, modeled as independent Gaussian currents.  $\psi_{\mbox{\tiny osc}}$ is the phase difference of the external oscillation of two sequential modules. $\phi_i$ is the phase difference, in module M1, between the oscillations peak and the first input. B) Map of the $f^{\;Best}_\gamma$ for a given set of $\phi_i$ and $\psi_{\mbox{\tiny osc}}$ when the oscillation is in the theta range $f_{\theta} = 8 Hz$. Blank area represents the condition where the four stimulus were not correctly allocated to the four modules. C) $f^{\;Best}_\gamma$ dependency on $\psi_{osc}$ for three values of $\phi_i$. Shaded areas show the upper and lower values of $f_\gamma$ that loaded A,B,C and D in M1,M2,M3 and M4.} 
	\label{fig23} 
	\end{centering}
\end{figure}

In order to better understand the properties of the model after the introduction of the modular structure, we address the question of if there is an optimal delay between stimuli (or an optimal presentation frequency) for the correct allocation of items into different modules. 
We imagine four items (A through D) being presented to be held in memory at a rate $f_\gamma$. This is translated in terms of a sequence of  Gaussian input  currents to the network (Fig. \ref{fig22} top). The Gaussian shapes of the inputs account for a certain degree of asynchrony in the upstream network representing the items and as a whole the input  resembles a bout of gamma oscillations of frequency $f_\gamma$ (Fig. \ref{fig23}A). The informational inputs are fed synchronously to all modules of the network, but they reach each of the modules on a different phase of the unspecific oscillatory input due its propagation speed ($v_{\mbox{\tiny osc}}$). The phase difference between consecutive modules is  $\psi_{\mbox{\tiny osc}}=2\pi f_{\mbox{\tiny osc}}d/ v_{\mbox{\tiny osc}} $  where $d$ is the physical distance between the center of two consecutive modules and $f_{\mbox{\tiny osc}}$ is the oscillatory frequency, and we make the simplifying assumption that all neurons in a given module are under the same phase. We define the input phase $\phi_{p,m}$ of the $p^{th}$ item in the module $m$ as the phase difference between the positive peak of the oscillation in that module and the item input time $$
\phi_{p,m}  = \phi_i - (p-1) \phi_\gamma + (m-1) \psi_{osc} 
$$
where $ \phi_\gamma =2\pi  f_{\mbox{\tiny osc}} / f_\gamma $ and  $\phi_i$
is the input phase of the first item in the first module (see Fig. \ref{fig23}A). The input phase is positive if the stimulus anticipates the peak. Therefore, to evaluate the success of the load operation in different circumstances, three parameters are available: the items presentation frequency $f_\gamma$, the phase difference between modules $\psi_{\mbox{\tiny osc}}$ and the input phase of the first item in the first module $\phi_i$.

We consider that the load operation is successful if 
items presented in a sequence are distributed in a orderly fashion among the different modules. The easiest of possible orders is to have each item stored in a different module (Fig. \ref{fig22}) so that the first item is represented in the first module, the second in the second, and so forth. The success of the load operation, therefore, depends on an optimal temporal match between the timings of the item inputs and the windows of opportunity  defined by the positive phases of the oscillatory drive in the different modules. This involves adjusting the $f_\gamma$, $\psi_{\mbox{\tiny osc}}$ and $\phi_i$ parameters. We devised a criterion that optimizes $f_\gamma$ given a choice of $\psi_{\mbox{\tiny osc}}$ and $\phi_i$. Figure \ref{fig23}B shows a map where the color indicates the optimal values of  $f^{\mbox{\tiny Best}}_\gamma$, and where the blank area is the condition where the four stimuli are never allocated correctly (see Materials and Methods). Figure \ref{fig23}C has the same information of \ref{fig23}B, but represented as different curves of ($f^{\mbox{\tiny Best}}_\gamma \times \psi_{\mbox{\tiny osc}}$) for each $\phi_i$ condition.

\begin{figure}[ht]
\begin{centering}
	\hspace{0.2cm}\includegraphics[width=0.8\linewidth]{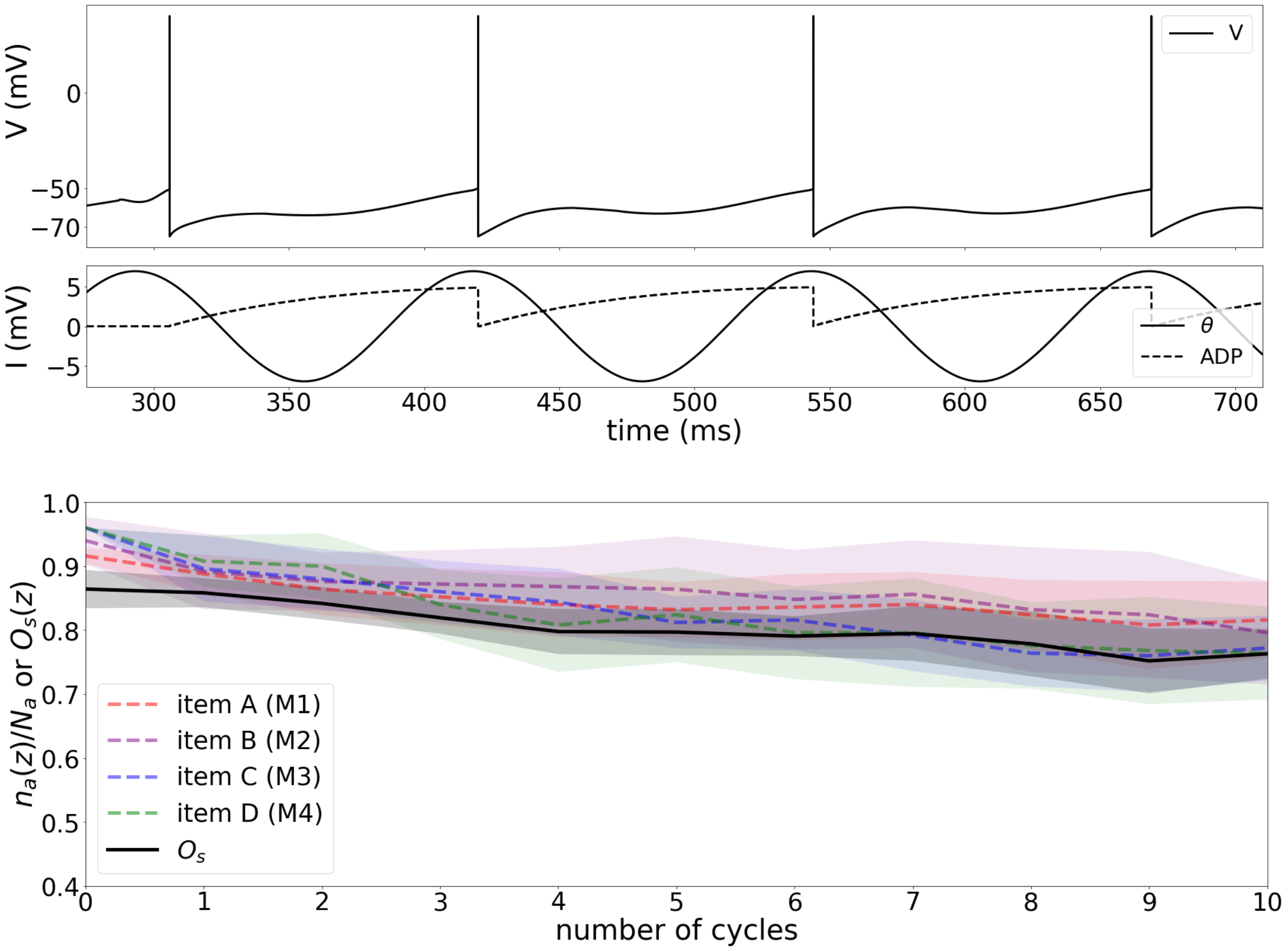} 
	\caption{Memory maintenance. A) Basic mechanism for maintenance at the neuronal level: the sum of the oscillatory theta input and the intrinsic afterdepolarization current is sufficient for a cyclic reactivation of the neurons once stimulated. B) Four items, one allocated to each module, are stably maintained in time for several theta cycles. The y-axis accounts for the fraction of active neurons of each item that during each cycle ($n_a(z)/N_a$, a $\in$ [A,B,C,D]) and the $O_s(z)$ parameter.  } 
	\label{fig24}
\end{centering}
\end{figure}

The next issue is how well the network maintains information stored through time. In the Lisman-Idiart model for multi-item working memory,  the combination of the oscillatory current in theta frequency and the membrane after-depolarization current creates an state of neuronal cyclic reactivation (figure \ref{fig24}A), therefore the same neurons activated by a stimuli will be perpetually reactivated. 

We consider that the maintenance operation is successfully accomplished if 
three requirements are met:
i) all the neurons representing an item are firing with a good degree of synchrony during memory maintenance ii)
each item is stored in a different module so neurons from different items fire asynchronously, iii) memory is reactivated in the first  cycle of the unspecific oscillation. Using these criteria, we devised an order parameter ($O_s(z)$) that measures the quality of the memory load and maintenance at a given cycle (z) (See Materials and Methods).  Higher values of $O_s(z)$ indicate better storage.Figure \ref{fig24}B shows that the four items correctly allocated are stably maintained through time by four different modules. 

\begin{figure}[ht!]
\begin{centering}
	\hspace{0.2cm}\includegraphics[width=0.65
	\linewidth]{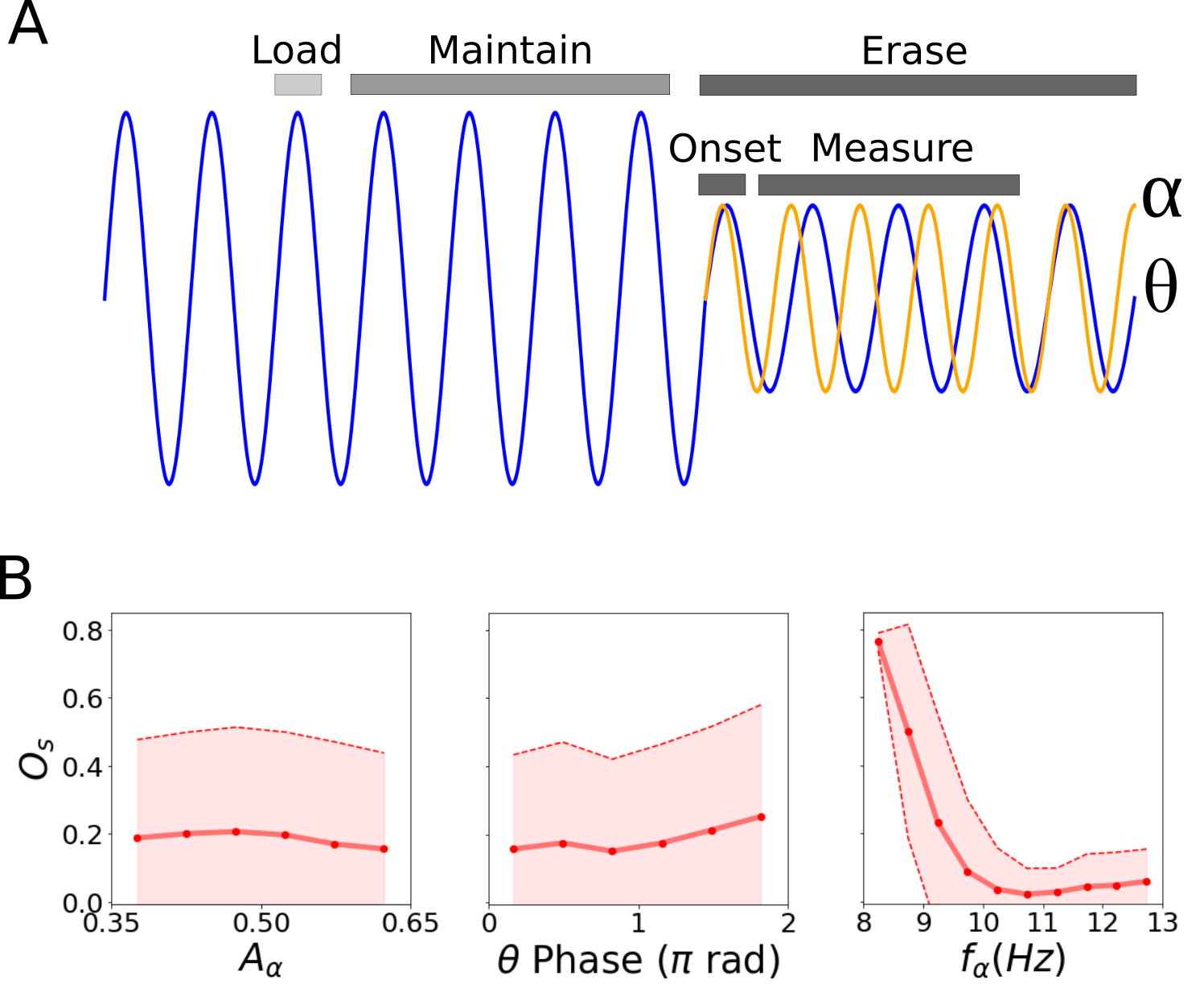} 
	\caption{Memory erasing. A) Simulation's protocol: Load - Information is loaded in the network; Maintain - Information is maintained by reactivation for four oscillatory cycles; Erase - The onset of alpha oscillations occurs and storage measures are made on the next three cycles. B) Multi-item working memory performance measured by the $O_s$ parameter during the first three theta cycles (1',2' and 3') after the onset of alpha. The plots show dependency with the amplitude of alpha, the onset theta phase and the alpha frequency. Thick red lines shows the binned mean, where the markers are the bin's center and the shadowed area is the standard deviation.} 
	\label{fig25} 
	\end{centering}
\end{figure}

While the power of theta oscillations is correlated with the maintenance of working memory, the power of alpha is thought to actively inhibit irrelevant information on the same tasks. We propose that a possible mechanism for this inhibitory effect is the interference between existing oscillations in the alpha e theta range in the same network.   
We call it Oscillatory Interference Hypothesis. We consider that the combination of alpha and theta frequencies produce a beat, and the cyclic reactivation of the firing patters proposed in Lisman-Idiart ends up impaired and canceled,  resulting in a {\bf erase} operation on the working memory buffer (fig \ref{fig25}A). Using the parameter $O_s(z)$ for each theta cycle, we show that the inhibitory role of alpha was independent of the amplitude and the initial onset, but dependent on the specific frequency  within the alpha band. Figure \ref{fig25}B shows the distribution of the average values of $O_s$  within the first three cycles after the onset of alpha (cycles 1', 2' and 3') for similar amplitudes of alpha and theta, initial theta phase between 0 and 2$\pi$, and alpha frequencies in the alpha band (8-13Hz). Thick red lines show the binned mean, where the markers are the bin's center and the shadowed area is the standard deviation.

\section{Discussion}
We proposed an adaptation of the Lisman-Idiart model where the neurons are structured in spatial modules and the oscillations are traveling waves. The main reasons for this are (i) an increase in the robustness of the model and (ii) test if the model is compatible with those properties. When proposed for the first time, the model used to segregate memories through fast-response inhibition, with sequential delays of the firing reactivations \cite{LismanIdiart_1995}. For ranging parameters, the model becomes instable and syncronizations between items occurs easily. The travelling wave solve this problem, segregating the oscillatory drive by a phase constant. Recent evidences suggesting that oscillations in the hippocampus and neocortex are traveling waves \cite{theta_alpha_cortex, theta_hip,theta_hip2} also encouraged us to incorporate this change in the model. The idea of a modular brain, where neurons connect together forming micro local networks, has being accepted over a while with new measures of brain connectivity. Our changes in the oscillatory activity lead us to think in space, and inevitable the multiple stored items should not only have temporal but also an spatial segregation. On the other hand, fixing memories positions in certain modules would not be compatible with the flexibility of changing the order of items. So we proposed multiple stem modules\footnote{As a stem cells analogy.}, with the capacity of coding any and all information. This approach also solves the problem of overlapping memories, when neurons can code for two different items. In short, two sequential coded items always will be in different modules, by this opens the possibility for overlapping representations of items.

First, we show that even if four stimulus are given equally to four modules, the excitatory dynamic created by an theta travelling wave allows the correct allocation of one memory in each module. The best frequency of stimulation seems to be related to the speed of the traveling wave, binding, once more, the theta and the gamma oscillations. The best gamma frequency is approximately the one in which the gamma period matches the difference in time between the phases of theta oscillations of two sequential modules, being smaller as the global inhibition increases. Recent studies investigating theta and alpha oscillations as travelling waves indicate a phase difference of 2$\pi$ rad/cm in the rat hippocampus \cite{theta_w}, $0.1 rad/cm$ in the human hippocampus \cite{theta_hip} and $0.3-1.25 rad/cm$ in the human neocortex \cite{theta_alpha_cortex}, suggesting that our model would fit better on a human neocortical surface circuit, with a spatial scale of centimeters in and between modules. Our model indicates that there is a wide range of parameter allowing the theta and gamma oscillations to create the multi-item working memory storage. The blank area of the map of figure 3b represents a more complex condition where more than one item is coded on the same module or the same item is coded by two different modules. Since every module has all the features of the Lisman-Idiart model, it could operate as local short-term buffers in case there is no interaction between modules. One possibility is that, regarding evidences about the role of the direction of propagation of the travelling waves in working memory \cite{theta_alpha_cortex}, the coherence of the propagation of the travelling waves could recruit more or less modules, turning on the spatial features or maintaining just the local ones.

We use the same maintenance mechanism proposed on the Lisman-Idiart model, where stimulated neurons presents an afterdepolarization currents and the sum with the oscillatory theta input allows a cyclic reactivation. So, the same neurons once stimulated will fire again on each theta cycle. The same patterns activated by the stimuli A,B,C and D are repeated on each cycle for the modules M1, M2, M3 and M4.

The Oscillatory Interference Hypothesis showed to be a plausible mechanism for the inhibitory role of the alpha oscillations.Considering a condition where the alpha and theta oscillations compete for power (similar amplitudes), Alpha was able to disrupt the working memory performance independently of its amplitude. The initial theta phase onset of the increase of alpha also not contributed for the inhibitory effect. The frequency appears to act as an threshold, and for a theta with frequency 8Hz, values of alpha frequency above 10Hz  effectively erased the  working memory buffer. The measures taken used the combination of three cycles to assure the time needed for the system to show long lasting and stable behaviour.

Having in mind the nature of beats, the key mechanism by which alpha is actively inhibiting the working memory maintenance is by producing a long lasting period of low excitability (Fig. 6A). Using an condition where both alpha and theta are synchronized at phase = 0 during the alpha's onset, we can see which values of alpha (and higher frequency bands) could produce a beat where the valley coincides with window of highest ADP amplitude ($I_{ADP} > 0.7A_{ADP}$), for theta = 8Hz (Fig 6B). This could explain the alpha frequency dependency found during our simulations. For other values of $f_\theta$, Fig. 6C shows the dependence between the average value for alpha ($\bar{f}_\alpha = f_{\alpha}^{max} - f_{\alpha}^{min}$) and theta frequency .

\begin{figure}[ht]
\begin{centering}
	\hspace{0.2cm}\includegraphics[width=0.6
	\linewidth]{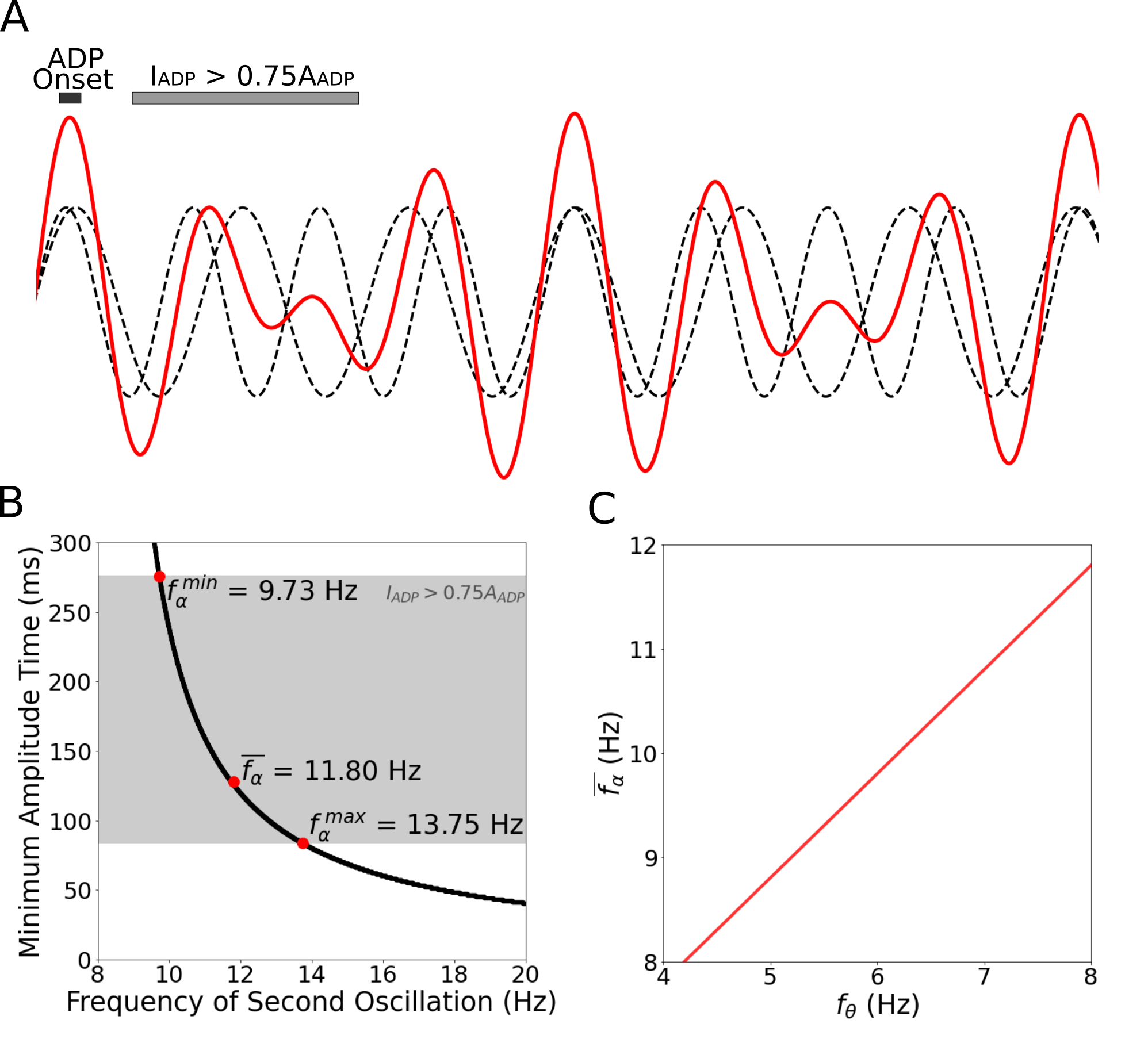} 
	\caption{Alpha inhibitory mechanism. A) Beat, in red, produced by an theta (8Hz) and alpha oscillation (10 Hz), dashed black lines, starting synchronized with phase = 0. B)  Values of alpha (and higher frequency bands) could produce a beat where the valley coincides with the next two theta peaks, for theta = 8Hz. C) Mean possible alpha ($\bar{f}_\alpha = f_{\alpha}^{max} - f_{\alpha}^{min}$)for other values of $f_{\theta}.$ } 
	\label{fig6} 
	\end{centering}
\end{figure}

\section{Conclusion}
In this letter, we discuss a possible mechanism behind the blockage effect that oscillations in the alpha range appear to be in cognitive tasks that demand that the subjects disregard parts of the external stimuli.
We do it in the light of the theta-gamma model proposed by Lisman-Idiart. In this model memory maintenance depends on two factors, an intrinsic excitability caused by recent activity that tags neurons associated with a given memory and an oscillatory input, in the theta range, that is sub-threshold for neurons that are not part of the memory but can drive the tagged neurons back to activity keeping them in oscillatory persistent activity, since there is a resonance effect between the time course of the excitability and the oscillatory input. According to the logic of the model, memory erasure could be accomplished by either eliminating the neural intrinsic response to firing or by disrupting the oscillatory input that refresh the memory.
We propose that an efficient way to disrupt the oscillation is by gating to the circuit an oscillation in the alpha range that will be superimposed to the existing theta oscillation causing amplitude modulations with the exact time scale as to prevent the tagged neurons of refreshing their excitability.
A possible criticism to that is to argue that the same effect can be easily accomplished by reducing the power of the theta oscillations. Although it is a valid statement here we subscribe to the view that oscillations are natural attractors for the biological neural networks and preventing then may be more energetically costly than just combining them.\\

\section{Materials and Methods}
NETWORK\\
We consider a network composed by $N$ neurons, which $N_{ex}$ are principal excitatory and $N_{inh}$ are inhibitory interneurons. The network is divided in $M$ spatial modules,with $N_a/M$ neurons of each type, $a=ex,inh$. The neural connectivity depends on the modules the neurons belong, as well as the oscillatory inputs, with sequential oscillatory phase-differences producing the effect of a travelling wave. The stimulus input and the output occur simultaneously for all modules.\\
\vspace{0.6cm}

CONNECTIONS\\
The network follow the connectivity matrix of Supplementary Figure 1, where the strength is randomly defined in a uniform distribution between 0 and the connection type parameter $W_{type}$.

\begin{equation}
    W_{ab} \sim U(0, W_{type})
\end{equation}

Where $W_{ab}$ is the connection from the presynaptic neuron a to postsynaptic neuron and $U(0,W_{type})$ is a uniformly distributed random number between 0 and $W_{type}$. $W_{type}$ can be Global ($W_{EI}$, $W_{IE}$) or Modular ($W_{EE}$, $W_{EI}$, $W_{IE}$).
\vspace{0.6cm}

NEURONS\\
The neurons are modeled as current based integrate and fire, given by the equation
\begin{equation}
    \tau_m\frac{dV}{dt}= - (V-V_{r}) +\sum_i    I_i
\end{equation}
where $V$ is the membrane potential, $V_{r}$ is the resting potential, $\tau_m$ is the membrane time constant and the last term is the sum over the input currents. The potential is reset to a hiperpolarized value $V_{reset}$ after passing the firing threshold of $V_{threshold}$, and stay unable to fire again for refractory time of $t_{refractory}$.
\vspace{0.6cm}

CURRENTS\\
The total post-synaptic input received by a neuron i due to the firing of other neurons  in a given time t is
\begin{equation}
I^{ps}_{i}(t)= \sum_{j=1}^N W_{ij} \sum_{s=1}^{n_j(t)} P(t-t^{(s)}_j )  \;
\end{equation}
 where $W_{ij}$ is the synaptic weight, $n_j(t)$ is the number of spikes fired by neuron $j$ up to time $t$, and $t^{(s)}_j$ are the spike times,  and the individual input P is 
\begin{equation}
    P(t) = H(t) e^{-t/\tau_{psp}}
\end{equation}
where $H$ represent a Heaviside function.
The principal excitatory neurons have an afterdepolarization potential
that is reset for each new spike. 
\begin{equation}
I_i^{ADP}(t)= A_{ADP} \; \left( \frac{t-t_i^*}{\tau_{ADP}} \right) \; e^{-(t-t_i^*)/\tau_{ADP} + 1} \; H(t-t_i^*)
\end{equation}
where $t_i^* < t$ is the last spike of cell $i$ before time $t$.
 The informational stimulus are modeled as Gaussian pulses 
\begin{equation}
I^{inf}_i(t) = A_{inf} \;\delta_{i\in A} \; e^{-(t-t_A)^2/(2\sigma_\gamma^2)}
\end{equation}
where $A_{inf}$ is the input amplitude, $t_A$ the average time the stimulus was presented to the network and  $\delta_{i\in A}$ indicates that only neurons linked to the information pattern A receive the inputs.  The oscillatory inputs are given by the sinusoidal function
\begin{equation}
    I_i^{osc}(t)=A_{osc} \; \sin(2\pi f_{osc} t + (m_i-1)\psi_{osc})
\end{equation}
where $A_{osc}$ is the amplitude, $f_{osc}$ is the frequency,$\psi_{osc}$  is a phase-shift creating a travelling wave effect and $m_i=1,2,...,M$ is the module index of neuron $i$ . The oscillatory power is modeled as all external oscillations come from the same source, meaning that the total power is a constrain condition for the system.\\
\vspace{0.6cm}

NOISE\\
We introduce noise in the system as a variability in the firing threshold for each neuron
\begin{equation}
    V_{threshold,i} = -50 + \eta_i
\end{equation}
where  $$\eta_i \sim N(\mu_{noise},\sigma_{noise})$$ is a normally distributed random  variable  with mean $\mu_{noise}$ and standard deviation $\sigma_{noise}$ that is drawn independently for each neuron after each  new spike.
\vspace{0.6cm}

METRICS FOR LOADING PERFORMANCE\\
 In order to evaluate the best frequency  of stimuli presentation $f_\gamma$, for loading information into the network given the parameters ($\psi_{\mbox{\tiny osc}}$, $\phi_i$), we simulated the loading cycle (zero cycle) of our network varying $f_\gamma$  between 100Hz and 33.33Hz. We then counted the number of activated cells for each item for each module $n_{a,m}$ where $a\in [A,B,C,D]$ and $m \in [M1,M2,M3,M4]$. We consider that, in a first approximation, a frequency is suitable for loading if the order of the stimuli is preserved in the modules. In other words, if the first item is the winner (the most active) in the first module, the second item is the winner in the second module, and so on.   Mathematically a binary loading suitability can be written as
 \begin{equation}
 \ell(f|g) = \prod_{i} \prod_{j\ne i} H[  n_{i,i}(f) - g * n_{i,j}(f) ]
 \end{equation}
 where $H[\cdot]$ is the Heaviside function, the level $g>1$ is a parameter that controls how bigger the winners must be (for instance, $g=2$ indicates that the winner has to be at least twice the runner up),  and the indexes $i,j$ are numerical indexes representing items and modules considered here in equal number. 
 We assume that the most suitable frequency (the best $f_\gamma$)  for loading information for a given ($\psi_{\mbox{\tiny osc}}$, $\phi_i$)  as the average
 \begin{equation}
 f_\gamma^{\mbox{\tiny Best}}(\psi_\theta, \phi_i| g) = \frac{\sum_f \ell(f|g) f }{\sum_f \ell(f|g)} .
 \end{equation}
 When $\sum_f \ell(f|g)= 0$ there is no suitable frequency, at level $g$, and the result is represented by a blank in Fig. \ref{fig23}B.
\vspace{0.6cm}

METRICS FOR MAINTENANCE PERFORMANCE\\
 We developed an order parameter that can account for the system's storage properties. These are: neurons from a given stored item need to be synchronized, while stay asynchronized with neurons from other items. So, for the $z^{th}$ reactivation  cycle and a stored item $A$, the measure for the synchronization within a memory item is

  \begin{equation}
 O^{Syn}_{A}(z)  =  \frac{n_{A}(z)}{N_A}\; \left[ 1 - \left( \sqrt{2} \; \frac{\sigma_{A}(z)}{\Delta t}  \right)^{\beta_{s}}  \right]_{+}
 \end{equation}\\
 
 where $n_{A}(z)$ is the number of active neurons in the ensemble that represents item $A$ at the $z^{th}$ cycle, $N_A$ is the total number of neurons in the ensemble that represents item $A$, $\sigma_{A}(z)$ is the standard-deviation of the firing times of the $n_{A}(z)$ neurons, $\Delta t$ is the time between the reactivation of two sequential items,  $\beta_s$ is a control parameter that punishes the standard-deviation increase. The measure for asynchronization between two memory items $A$ and $B$, in the $z^{th}$ cycle  is
 
  \begin{equation}
  O^{Asyn}_{AB}(z) =  \phi \left( \frac{ |\langle t_{A}(z) \rangle   -   \langle t_{B}(z) \rangle |  }{\Delta t} \right)  
\end{equation}
where
$$ \langle t_{A}(z) \rangle  = \frac{1}{n_A(z)}  \sum_{i}  t_{A,i}(z) $$
with $t_{A,i}(z)$ the firing times for the individual neurons representing item $A$
and $$\;\;\;\;\; \phi (x) = \left\{ \begin{array}{cl}  0  & \mbox{if}  \;\;  x = 0   \\  x^{\beta_{a}}  & \mbox{if}  \;\;  x  \in (0,1)  \\   1  & \mbox{if}  \;\;  x \ge 1    \end{array}  \right.$$\\
where $\beta_a$ a parameter controlling  $\phi$'s non linearity.

Our order parameter therefore the multiplicative  combination of the average over items of both measures, in a given cycle $z$,
 \begin{equation}
 O_s(z) =  \left[  \frac{1}{M}  \sum_{A}  O^{Syn}_{A}(z) \right] \left[ \frac{2}{M(M-1)}  \sum_{A,B>A} O^{Asyn}_{AB}(z) \right] \;\; . 
 \end{equation}
 
\vspace{0.8cm}
BINNED STATISTICS\\
We used binned statistics in order to assess if the capacity of alpha's inhibitory role in disrupting the working memory depended on its amplitude, frequency and initial onset.\\

\vspace{0.6cm}

SIMULATIONS AND DATA ANALYSIS\\
We used Euler's method for solving numerically the neurons differential equations.  The simulations were written in C programming language  and data analysis and graphic production were made with Python.\\
\vspace{0.6cm}

PARAMETERS\\
Parameters are shown in table \ref{tab_b}.
\vspace{2cm}

\begin{table}[h!]
\centering
\begin{tabular}{lcl}\hline
  &Fixed Parameters&\\\hline

Network                 &  & $N$ = 500                   \\
                        &  & $N_{ex}$ = 400                   \\
                        &  & $N_{inh}$ = 100                  \\
                        &  & $M$ = 4                  \\
Neurons                 &  & $\tau_{me}$ = 15 ms\\
                        &  & $\tau_{mi}$ =  2 ms\\
                        &  & $\tau_{ADP}$ = 140 ms \\
                        &  & $A_{ADP}$ = 7 mV\\
                        &  & $V_{rest}$ = -60 mV\\
                        &  & $V_{thresh}$ = -50 mV\\
                        &  & $V_{reset}$ = -70 mV\\
                        &  & $T_{ref}$ = 3 ms\\
Psp current             &  & $\tau_{ps.e}$ = 1 ms\\
                        &  & $\tau_{ps.i}$ = 10 ms\\
Global Synaptic Weights &  & $\overline{W}_{EI}$ = 0.56\\
                        &  & $\overline{W}_{IE}$ = -0.056\\
Modular Synaptic Weights&  & $\overline{W}_{EE}$ = 0.35\\
                        &  & $\overline{W}_{EI}$ = 2.25\\
                        &  & $\overline{W}_{IE}$ =-0.4\\
$O_s$ Parameter            &  & $\beta_{s}$ = 1\\
                        &  & $\beta_{a}$ = 1\\
                        &  &$\Delta t$ = 20 ms\\
Others                  &  & $\sigma_{noise}$ = 0.5 mV\\
                        &  & $\mu_{noise}$ = 0 mV \\
                        &  &$\sigma_\gamma$= 4 ms\\
                        &  & $dt$ = 0.01 ms\\
\hline
  &Variable Parameters&\\\hline

Oscillations & & $f_{\theta}$ = 4-8 Hz \\
             & & $f_{\alpha}$ = 8-13 Hz \\
             & & $A_{\theta}$ = 0.35-0.65 max \\
             & & $A_{\alpha}$ = 0.35-0.65 max\\
             & & $\psi_{osc}$ = 0 - 1.2 rad/module \\ 
             & & $\phi_i$ = 0.2 - 0.8 rad\\
             & & $f_\gamma$ = 1000 - 33.33 Hz\\


\end{tabular}
\caption[]{Overview of parameters. Top: fixed parameters held constant throughout the simulations. Bottom: variable parameters uses more than one value or a range of values. }
\label{tab_b}
\end{table}




\newpage
\section{Supporting Information}
\subsection{Connectivity Matrix}
\begin{figure}[ht]
\renewcommand{\thefigure}{S1}
\begin{centering}
	\hspace{0.2cm}\includegraphics[width=0.6
	\linewidth]{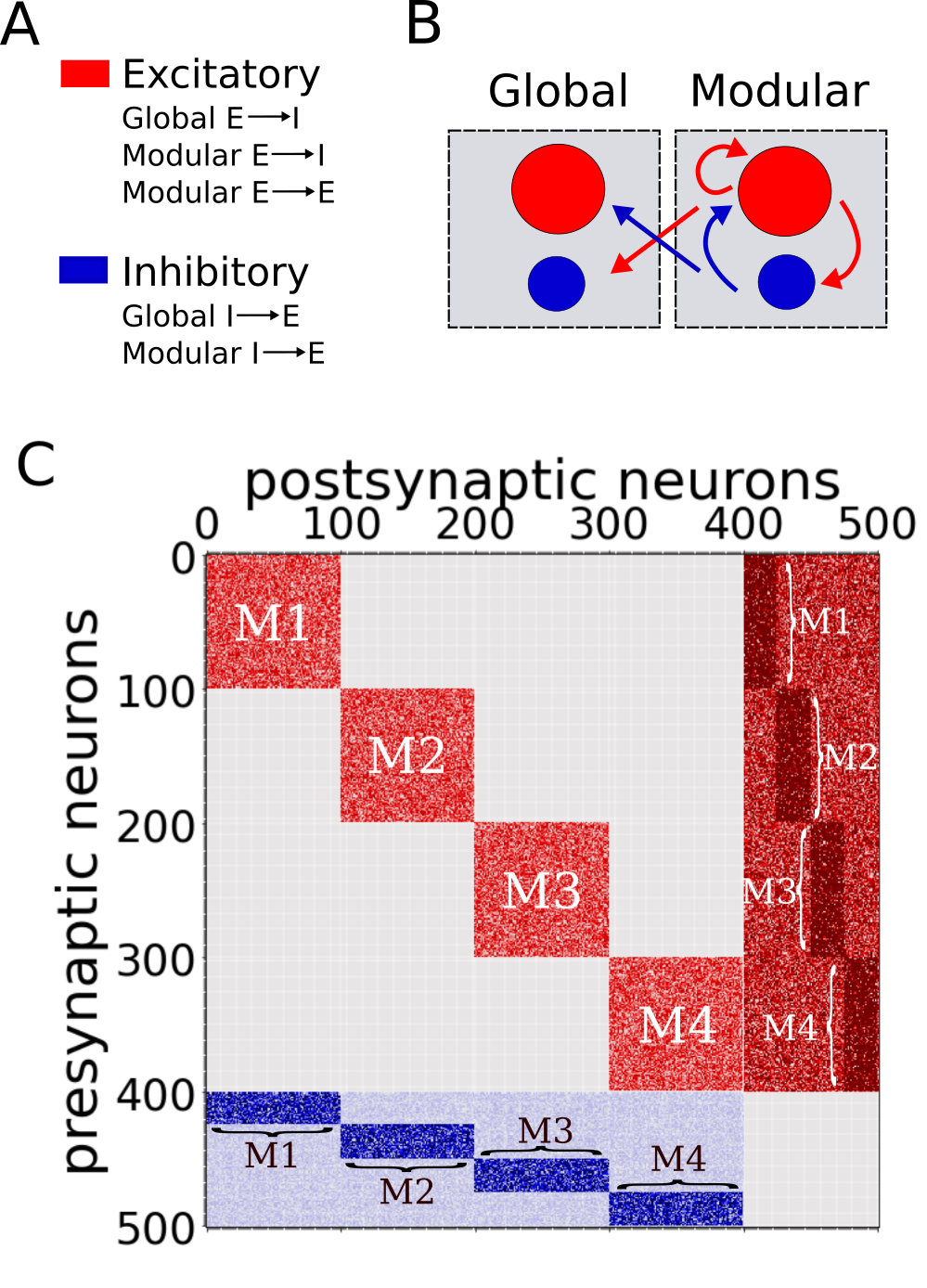} 
	\caption{Connectivity Specification. A) List of excitatory and inhibitory connections. B) Scheme of Global and Modular connections. C) Connectivity matrix for the network. The y-axis represent the presynaptic neurons and the x-axis the postsynaptic neurons.The excitatory and inhibitory neurons are grouped together for convenience.  } 
	\label{figS1} 
	\end{centering}
\end{figure}

 \newpage
 \subsection{Alpha Inhibition Experimental Data Points}
 \begin{figure}[ht]

 \renewcommand{\thefigure}{S2}
 \begin{centering}
 	\hspace{0.2cm}\includegraphics[width=0.8
 	\linewidth]{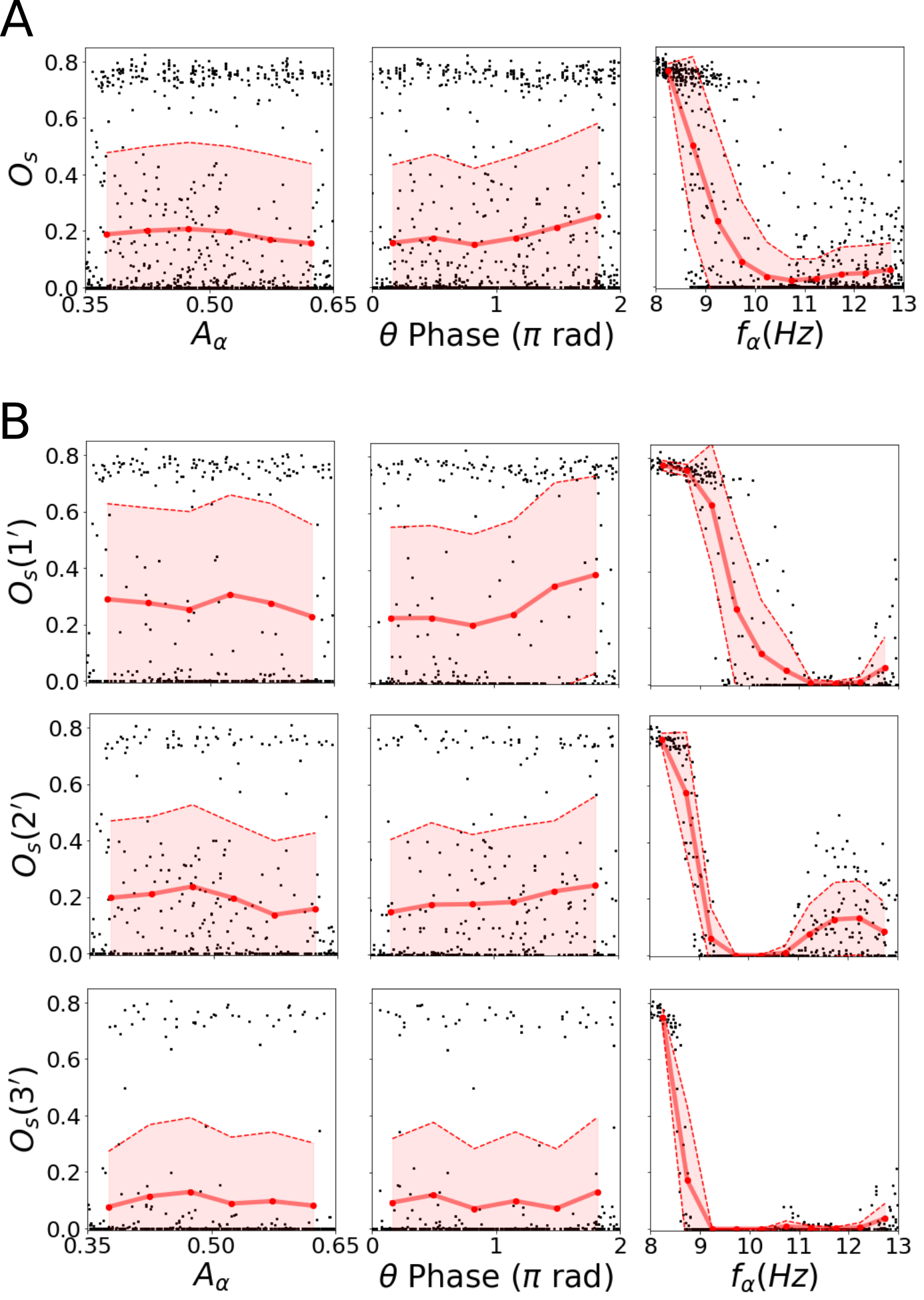} 
 	\caption{A) Data points for the experiments in Figure 5B. Black dots correspond to all measures taking during the three consecutive cycles after the onset of alpha. Binned statistics are shown in red, same as figure 5B. B) The same as A), but discriminating the cycles 1',2' and 3'.} 
 	\label{figS2} 
 	\end{centering}
 \end{figure}

\newpage

\section{Load Operation using Alpha insetad of Theta}
 \begin{figure}[ht]
 \renewcommand{\thefigure}{S3}
 \begin{centering}
 	\hspace{0.2cm}\includegraphics[width=0.8
 	\linewidth]{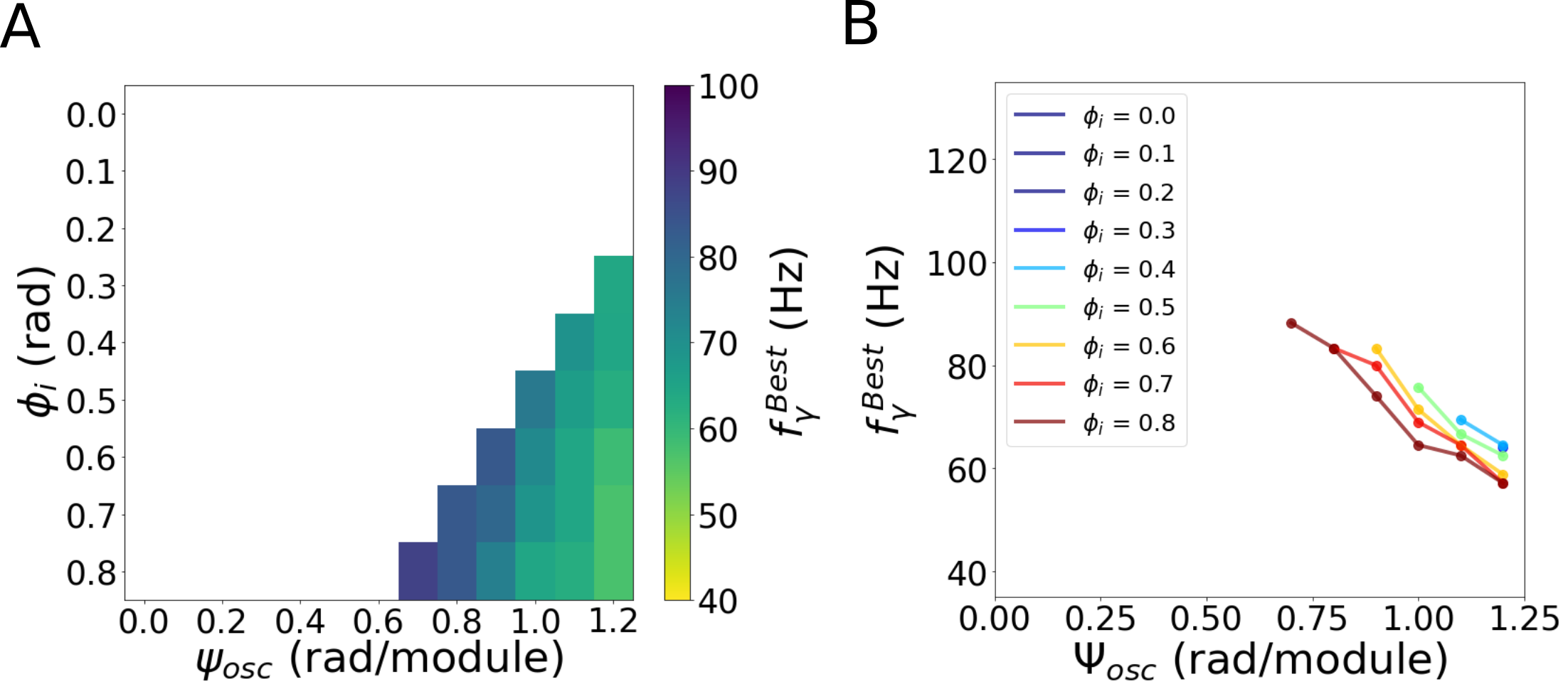} 
 	\caption{Load operation using alpha 12 Hz instead of theta 8 Hz. Similar plot as figure 3B and 3C. } 
 	\label{figS3} 
 	\end{centering}
 \end{figure}

 \newpage

\section{Additional Load Operation Plots}
 \begin{figure}[ht]
 \renewcommand{\thefigure}{S3}
 \begin{centering}
 	\hspace{0.2cm}\includegraphics[width=0.8
 	\linewidth]{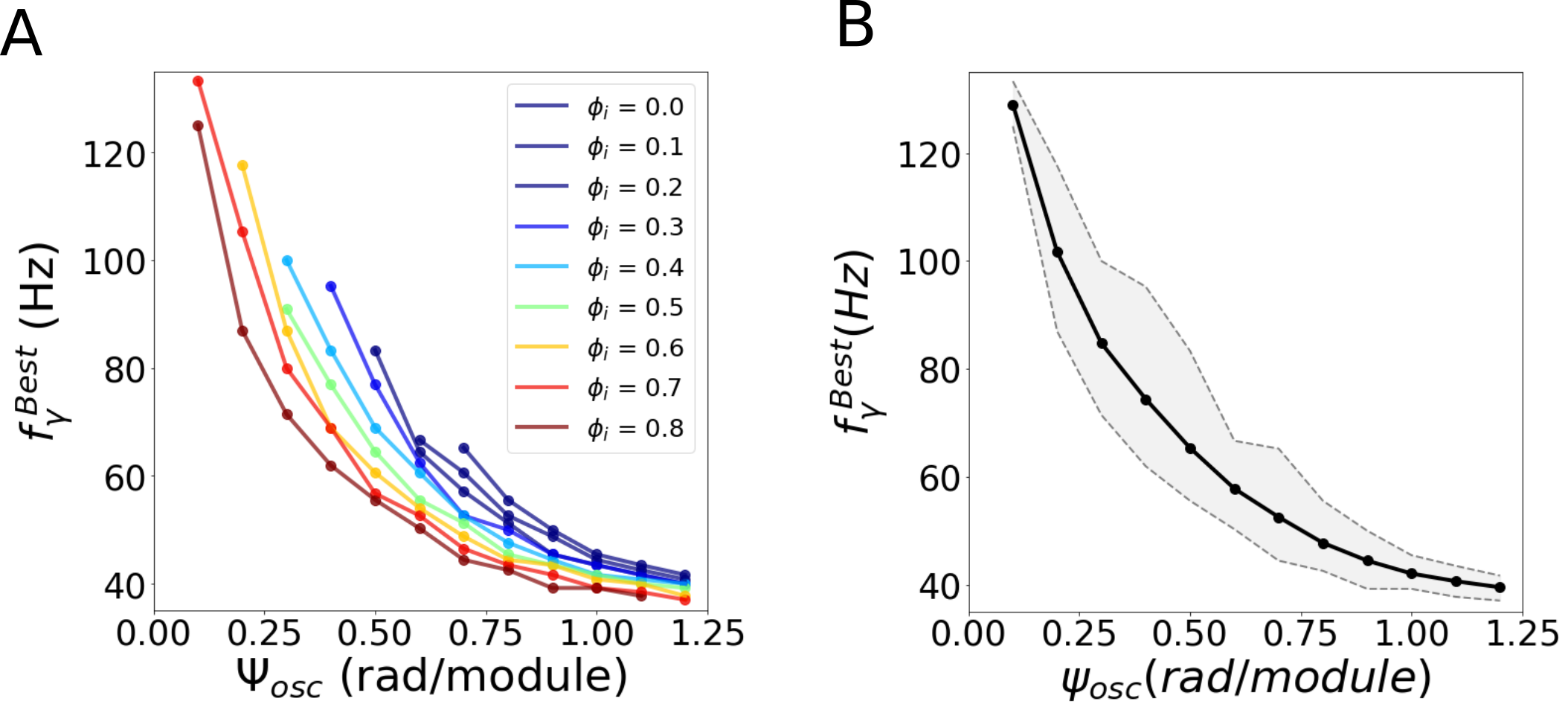} 
 	\caption{A) $f^{\;Best}_\gamma$ vs $\psi_{osc}$  for the complete set of $\phi_i$. B) Mean between $\phi_i$ conditions.} 
 	\label{figS4} 
 	\end{centering}
 \end{figure}




\end{document}